\documentclass[twocolumn]{emulateapj}

\usepackage{latexsym,amssymb}
\usepackage{amsmath}
\usepackage{natbib}
\usepackage{graphicx}
\usepackage{epstopdf}

\usepackage{xspace}
\usepackage{hyphenat}
\usepackage[colorlinks=true,linkcolor=blue,citecolor=blue]{hyperref}

\newcommand{\rt}{r_{\rm t}}

\newcommand{\FLASH}{{\tt FLASH}\xspace}
\newcommand{\GADGET}{{\tt GADGET-2}\xspace}
\newcommand{\Sim}{\mathord{\sim}}

\slugcomment{Draft \today}

\shorttitle{Turbovelocity Stars}

\shortauthors{Manukian et al.}

\begin{document}
\author{Haik Manukian\altaffilmark{1}, James Guillochon\altaffilmark{1}, Enrico Ramirez-Ruiz\altaffilmark{1} and Ryan M.\ O'Leary\altaffilmark{2,3}}
\altaffiltext{1}{Department of Astronomy and
  Astrophysics, University of California, Santa Cruz, CA 95064}
  \altaffiltext{2}{Department of Astronomy, University of California, Berkeley, CA 94720, USA}
  \altaffiltext{3}{Einstein Fellow}
   
\email{jfg@ucolick.org}

\title{Turbovelocity Stars: Kicks Resulting From the Tidal Disruption of Solitary Stars}
\begin{abstract} 
The centers of most known galaxies host supermassive black holes (SMBHs). In orbit around these black holes are a centrally-concentrated distribution of stars, both in single and in binary systems. Occasionally, these stars are perturbed onto orbits that bring them close to the SMBH. If the star is in a binary system, the three-body interaction with the SMBH can lead to large changes in orbital energy, depositing one of the two stars on a tightly-bound orbit, and its companion into a hyperbolic orbit that may escape the galaxy. In this {\it Letter}, we show that the disruption of {\it solitary} stars can also lead to large positive increases in orbital energy. The kick velocity depends on the amount of mass the star loses at pericenter, but not on the ratio of black hole to stellar mass, and are at most the star's own escape velocity. We find that these kicks are usually too small to result in the ejection of stars from the Milky Way, but can eject the stars from the black hole's sphere of influence, reducing their probability of being disrupted again. We estimate that $\Sim 10^5$ stars, $\Sim 1\%$ of all stars within 10 pc of the galactic center, are likely to have had mass removed by the central black hole through tidal interaction, and speculate that these ``turbovelocity'' stars will at first be redder, but eventually bluer, and always brighter than their unharrassed peers.
\end{abstract}

\keywords{black hole physics --- galaxy: center --- gravitation --- hydrodynamics --- methods: numerical --- stars: kinematics and dynamics}

\section{Introduction}

The orbits of stars in galactic nuclei are controlled by the combined influence of all other stars and the central supermassive black hole (SMBH). There is a chance that encounters between these stars can shift a star onto a nearly radial orbit which will bring it very close to the black hole \citep{Frank:1978wx}. When the distance of closest approach  $r_{\rm p}$ approaches $r_{\rm t}$, the star is distorted by tides, and the action of raising these tides excite normal modes of oscillation that transfer energy from the object's orbit to internal motions \citep{Press:1977ci}. For deeper encounters still, the tidal distortion can become so great that the star begins to lose mass, eventually being fully disrupted at a critical value of $\beta \equiv r_{\rm t}/r_{\rm p}$ \citep{Guillochon:2013jj}.

For dynamically relaxed stellar clusters surrounding SMBHs, the majority of stars that are placed on disruptive ``loss-cone'' orbits originate from the SMBH's sphere of influence \citep{Magorrian:1999vz}. As the velocity dispersion $\sigma$ at the SMBH's sphere of influence is smaller than the star's escape velocity $v_{\rm esc}$, even stars on initially hyperbolic trajectories can lose enough orbital energy during a tidal encounter to become bound to the SMBH. Once bound, these stars can have repeated interactions with the SMBH, which may eventually result in their destruction \citep{Antonini:2011ia}. Assuming that the star's orbit is not affected by other stars within the cluster, the continual injection of energy into the star can potentially produce a population of tidally-heated, tightly-bound stars \citep{Alexander:2003jr,Li:2013ij}. 

For tidal encounters in which the perturbed object loses no mass, the change in orbital energy is always negative, i.e. the star becomes progressively more bound to the black hole with each encounter. However, it has been found for sufficiently deep encounters that the asymmetry of the mass loss facilitates an energy exchange between the surviving object and the material that is removed from it at pericenter, conferring upon it a positive orbital energy \citep{Faber:2005be,Guillochon:2011be,Liu:2013er}.

In this {\it Letter}, we show through hydrodynamical simulations of stellar disruptions that stars that lose a large fraction of their own mass receive a velocity ``kick'' $v_{\rm kick}$ whose magnitude is {\it independent} of the mass ratio $q$, despite the fact that the asymmetry of the mass lost decreases with increasing $q$. Similar to what is found for moderate $q$ encounters, this kick results in a change in the star's velocity at infinity $v_{\infty}$ that can be as large as its own escape velocity $v_{\rm esc}$. As more massive stars have larger $v_{\rm esc}$, they will receive greater kicks. Through simple two- and three-body calculations, we compare the kicks received by solitary stars to the kicks received by stars through the disruption of binary systems. On average, we find that solitary kicks deposit the surviving \nohyphens{``turbovelocity''} stars (TVS) immediately beyond the sphere of influence $r_{\rm sph}$. As the relaxation time at $r_{\rm sph}$ is long compared to the age of the Milky Way \citep{Merritt:2010bz}, we anticipate that many stars within $r_{\rm sph}$ have had a strong tidal interaction with the Milky Way's SMBH, with a unique appearance that would likely make them distinguishable from unharrassed stars.

This {\it Letter} is structured as follows. In Section \ref{sec:method} we describe the setup of our hydrodynamical simulations. In Section \ref{sec:kicks} we explain how kicks arise, and why the magnitude of the kicks do not depend on $q$. In Section \ref{sec:discussion} we apply $v_{\rm kick}$ as determined by our simulations to predict the distribution of TVS within the Milky Way.

\section{Hydrodynamical Simulations  of  Partial  Disruptions Solar-type Stars}\label{sec:method}

\subsection{FLASH Simulations}
In a recent study \citep{Guillochon:2013jj}, we investigated the importance of the impact parameter and stellar structure in tidal disruptions of main-sequence (MS) stars by varying the impact parameter $\beta=r_{\rm p}/\rt$ and the polytropic index $\gamma$ with $q = 10^{6}$ and $e=1$. In this {\it Letter}, we present twelve different simulations of solar-type stars ($\gamma=4/3$) approaching SMBHs with varying $q=[10^3,10^6]$ and  impact parameters $\beta=[1.0,1.8]$, again along parabolic trajectories. The parameter space in $\beta$ and $q$ was chosen such that a surviving core always remains. As in \citeauthor{Guillochon:2013jj}, our simulations  are performed in \FLASH, an adaptive-mesh grid-based hydrodynamics code which includes self-gravity. Our method is identical to that of \citet{Guillochon:2011be} and \citet{Guillochon:2013jj}, with the exception that we calculate the star's self-gravity using a higher-order multipole expansion of the fluid ($l_{\max} = 40$).

\subsection{SPH Verification}
To verify our results, two simulations were also run using the SPH code \GADGET \citep{Springel:2005cz}. The initial stars used were the same as in the \FLASH simulations, and were initialized with $10^4$ equal mass particles. The stars were then placed on parabolic orbits around a SMBH with $q=4\times 10^6$. Convergence was tested by increasing the integrator accuracy of the simulations, we found that $v_{\infty}$ had converged with $\sim 10$\% uncertainty. Within this error, the result of the \GADGET simulations are in agreement with the \FLASH simulations.

\subsection{Post-Disruption Orbits and Comparison to Binary Disruptions}
To determine the observed velocity distribution of post-disruption stars, two orbit integrator methods were used to construct a Monte Carlo ensemble. The first integrator we use is based on the {\it Projection} solver within the {\tt Mathematica} software suite, with the orbital energies and angular momenta of all bodies in the system taken to be invariants \citep{Hairer:2000io}. This integrator was used to simulate the orbits of both solitary and binaries at the time of disruption. For solitary disruptions, the change in orbital energy $\Delta \epsilon_{\rm orb}$ calculated from our hydrodynamical simulations is applied instantaneously at pericenter. Afterwards, a simpler integrator using the explicit modified midpoint method is used to evolve the stellar orbits within the combined potential of the black hole (with assumed mass $M_{\rm h} = 4 \times 10^{6} M_{\odot}$) and the galaxy, where the stellar density $\rho \propto r^{-7/4}$ \citep{Bahcall:1976kk} interior to the core radius $a_{\rm c} = 8$ pc, and the function defined in \citet{Bromley:2006fz} exterior to $a_{\rm c}$,
\begin{equation}
\rho(r) = \frac{\rho_{0}}{1 + \left(r/a_{\rm c}\right)^{2}},
\end{equation}
where $r$ is the distance to the black hole and $\rho_{0} = 1.27 \times 10^{4} M_{\odot}$ is the central density. 

\begin{figure}[t]
\centering\includegraphics[width=\linewidth,clip=true]{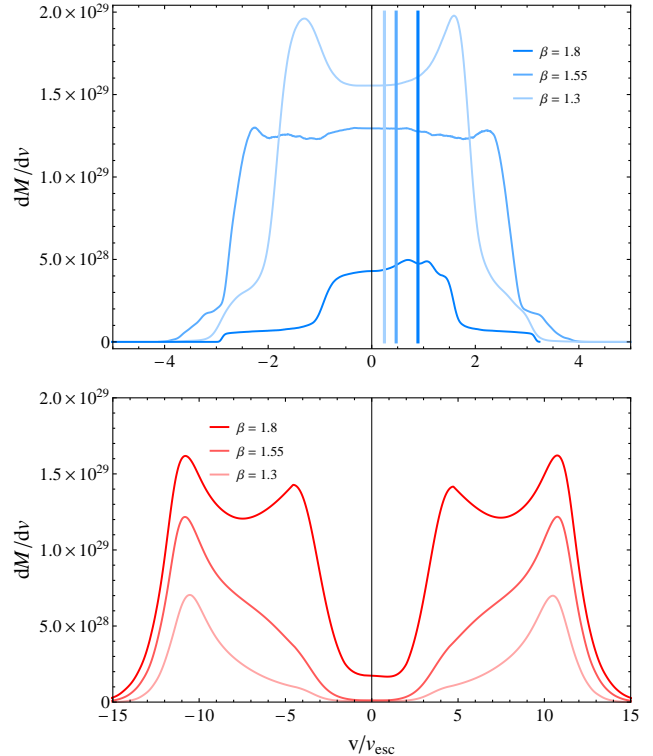}
\caption{Distributions of $dM/dv \equiv v(dM/dE)$ at the end of three different simulations for the material that is bound and unbound to the surviving core, top and bottom panels respectively, where thinner lines correspond to lower values of $\beta$.}
\label{fig:dmde}
\end{figure}

\begin{figure*}[t]
\centering\includegraphics[width=\linewidth]{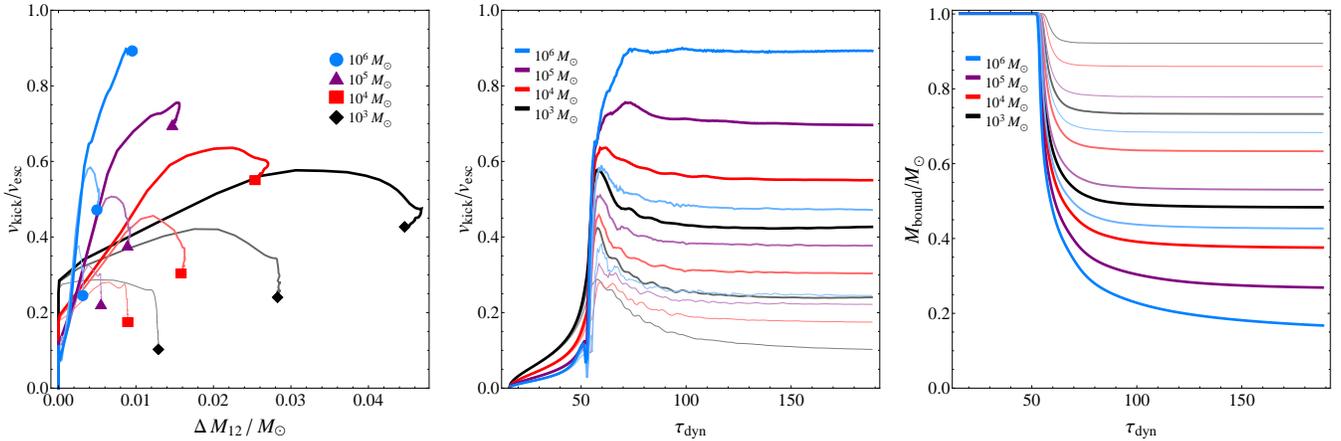}
\caption{In the left panel we show the convergence of $v_{\rm kick}$ vs.$\Delta M_{12}$ for all the simulations, where the lighter lines signify lower values of $\beta$. They converge to a value after the disruption event.  The middle panel shows $v_{\rm kick}$ convergence, where the lighter (and thinner) lines signify lower $\beta$ runs. And finally the plot on the right shows the final converged values of $M_{\textrm{bound}}$.}
\label{fig:convergence}
\end{figure*}

\begin{figure*}[t]
\centering\includegraphics[width=\linewidth,clip=true]{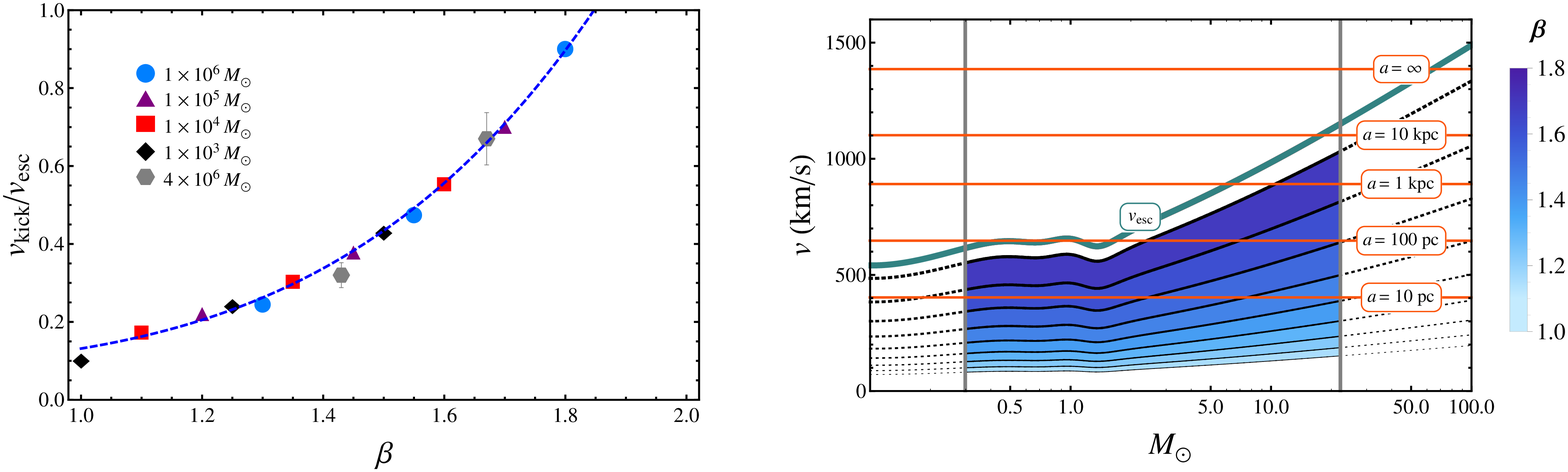}
\caption{The left panel shows $v_{\rm kick}$ delivered to the surviving core after disruption. A curve of the form $v_{\rm kick}/v_{\rm esc} = a + b \beta^{c}$ provides a reasonable description of the numerical results with $a = 0.0745$, $b = 0.0571$ and  $c = 4.539$ (blue dashed line). The right panel shows how $v_{\rm kick}$ at different $\beta$ scales with the mass of the pre-disruption star. The shaded area between the two mass limits encompasses stellar objects likely to be well-represented by a $\gamma = 4/3$ polytrope (stars whose convective regions are less than 50\% of their mass). The horizontal orange lines show the velocities required for stars to be placed on orbits of a given $a$.}
\label{fig:kicks}
\end{figure*}

\section{Asymmetric Mass Loss and Ensuing Kicks}\label{sec:kicks}
In a partial disruption event, the star sheds mass near the inner and outer Lagrange points L1 and L2, and forms two tails with mass $M_{{\rm L1}}$ and $M_{{\rm L2}}$. The mass loss is not symmetric; the difference in these two masses, $\Delta M_{12} \equiv M_{\rm L1} - M_{\rm L2}$ increases with smaller $q$. The asymmetry in the two tails becomes more evident with an increasing difference in the tidal field across the star as the mass ratio tends to unity. This asymmetry in mass loss is what induces the kick.

\subsection{Measuring Kick Velocities}
To obtain $v_{\rm kick}$ for the surviving stellar remnant from our simulations, we calculate the self-bound mass $M_{\rm bound}$ as well as the shift in energy of its centroid, $\Delta \epsilon_{\rm kick}$, as a function of time. From our simulations, the binding energy of material to the black hole $dM/dE$ for both the bound and unbound tails and the self-bound material can be readily estimated \citep{Guillochon:2013jj}. In Figure \ref{fig:dmde}, we show distribution of mass as a function of velocity $v$ for material that is bound and unbound to the surviving core at the end ($t_{\rm end} = 3 \times 10^{5}$ s) of our $q = 10^6$ simulations. A shift in the centroid of the self-bound material distribution is clearly seen, whose magnitude grows with increasing $\beta$.

The specific orbital energy $\epsilon$ of the surviving core is calculated by performing a mass weighted average
\begin{equation}
\epsilon= \frac{\sum_{i}m_i d\epsilon_i}{\sum_i m_i}
\end{equation}
and the difference between the initial and final $\epsilon$ is computed after $\epsilon$ has converged,
\begin{equation}
\Delta \epsilon_{\rm{kick}} = \epsilon(t_{\rm end}) - \epsilon(0).
\end{equation}
The kick velocity is then simply given by
\begin{equation}
v_{\rm{kick}} = \sqrt{2 \Delta \epsilon_{\rm{kick}}}.
\end{equation}
Figure \ref{fig:convergence} illustrates how quickly $M_{\rm bound}$ and $v_{\rm{kick}}$ converge. These kicks are at most $v_{\rm esc}$ and increase with the amount of stellar mass removed in the encounter.

\subsection{The Dependence of the Kick on $\beta$ and $q$}
We find, somewhat surprisingly, that $v_{\rm kick}$ is nearly independent of $q$ and depends on $\beta$ alone for a wide range of $q$ (Figure \ref{fig:kicks}). The majority of the asymmetry in the tidal response of the star originates from the excitation of the $l = 3$ mode, the most-significantly excited asymmetric mode \citep{Cheng:2013cm}. Because the asymmetry in the tidal force decreases with increasing $q$ \citep{Guillochon:2011be}, $\Delta M_{12}$ also decreases with $q$. However, the velocity of the star at pericenter $v_{\rm p}$ increases with $q$ for fixed $\beta$. In this section we show that the combination of these two effects, which together determine $v_{\rm kick}$, eliminates the $q$-dependence of the mass loss induced by excitation of the $l = 3$ mode.

When a star is tidally perturbed, its orbital energy will change, a well-known result in instances where the star loses no mass, as energy is transferred to oscillatory modes within the star \citep{Press:1977ci}. The linearized approximation of the change in $\epsilon$ for an initially non-oscillating star is
\begin{equation}
\Delta \epsilon_{\rm osc} = -q^{2} \frac{G M_{\ast}}{R_{\ast}} \left[\left(\frac{R_{\ast}}{r_{\rm p}}\right)^{6}T_{2} + \left(\frac{R_{\ast}}{r_{\rm p}}\right)^{8}T_{3}\right],\label{eq:deleosc}
\end{equation}
where $T_{2}$ and $T_{3}$ are functions that solely depend on $\beta$. If the star is oscillating, usually as a result of previous encounter(s) with its perturber \citep{Mardling:1995hx,Guillochon:2011be}, the signs of both terms in the square brackets in the above expression can be negative, enabling $\Delta \epsilon_{\rm osc}$ to be greater than zero. However, the final orbital energy $\epsilon_{\rm f}$ of the object can never be greater than the initial energy $\epsilon_{0}$ the object had before encountering the perturber for the first time, and thus a star that is initially bound to the black hole would always be bound to the black hole, even if $\Delta \epsilon$ is positive for a single orbit.

\begin{figure}[t]
\centering\includegraphics[width=\linewidth,clip=true]{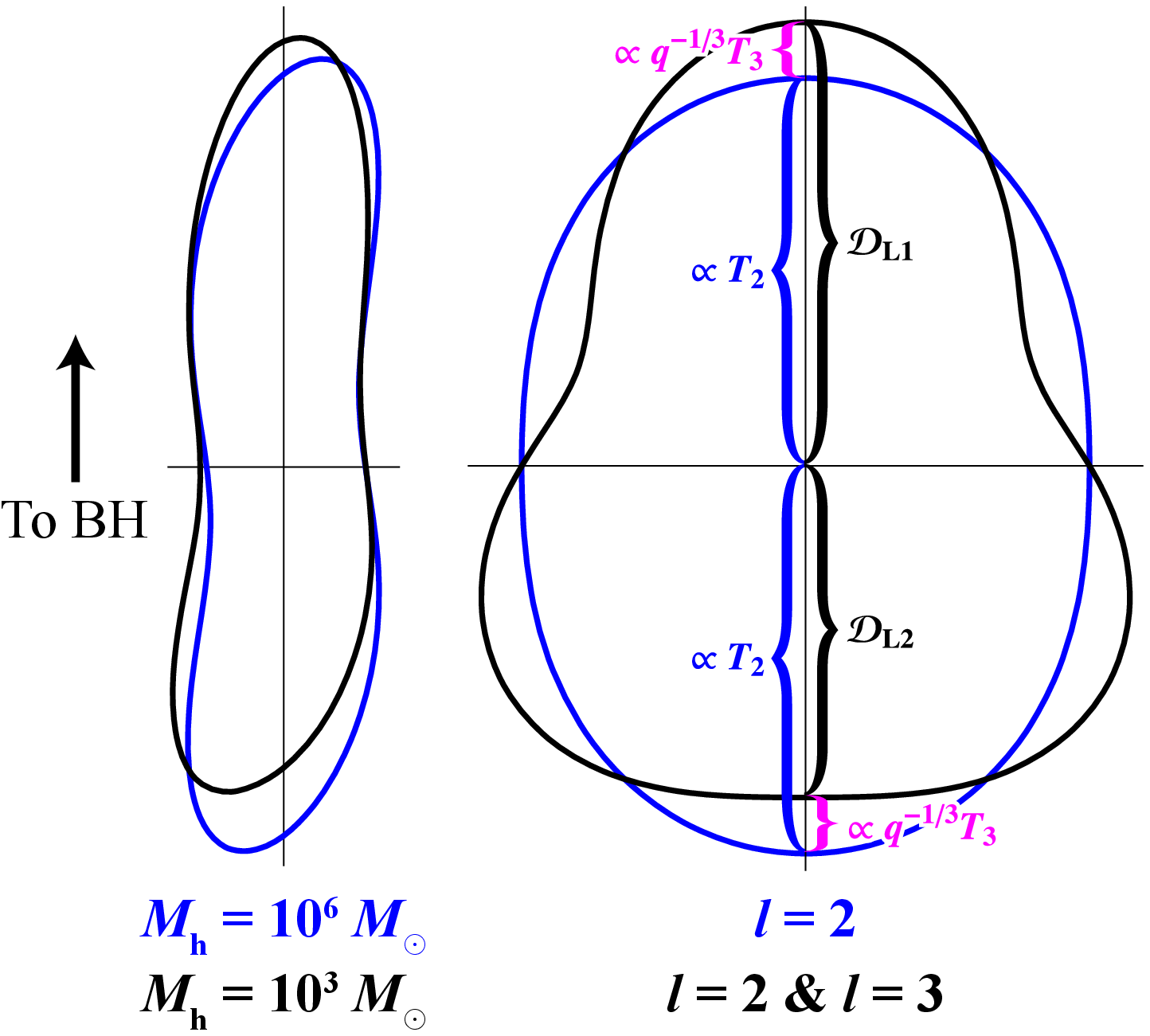}
\caption{Here we compare two contours of constant density $\rho$ = 1 g cm$^{-3}$ taken from a $M_{\rm h} = 10^6 M_{\odot}$ (blue) and $M_{\rm h} = 10^3 M_{\odot}$ (black) simulation, both with $\beta = 1.75$, to the functional forms of the $l = 2$ (blue) and $l = 2$ plus $l = 3$ (black) spherical harmonics.}
\label{fig:contours}
\end{figure}

\begin{figure*}[t!]
\centering\includegraphics[width=\linewidth,clip=true]{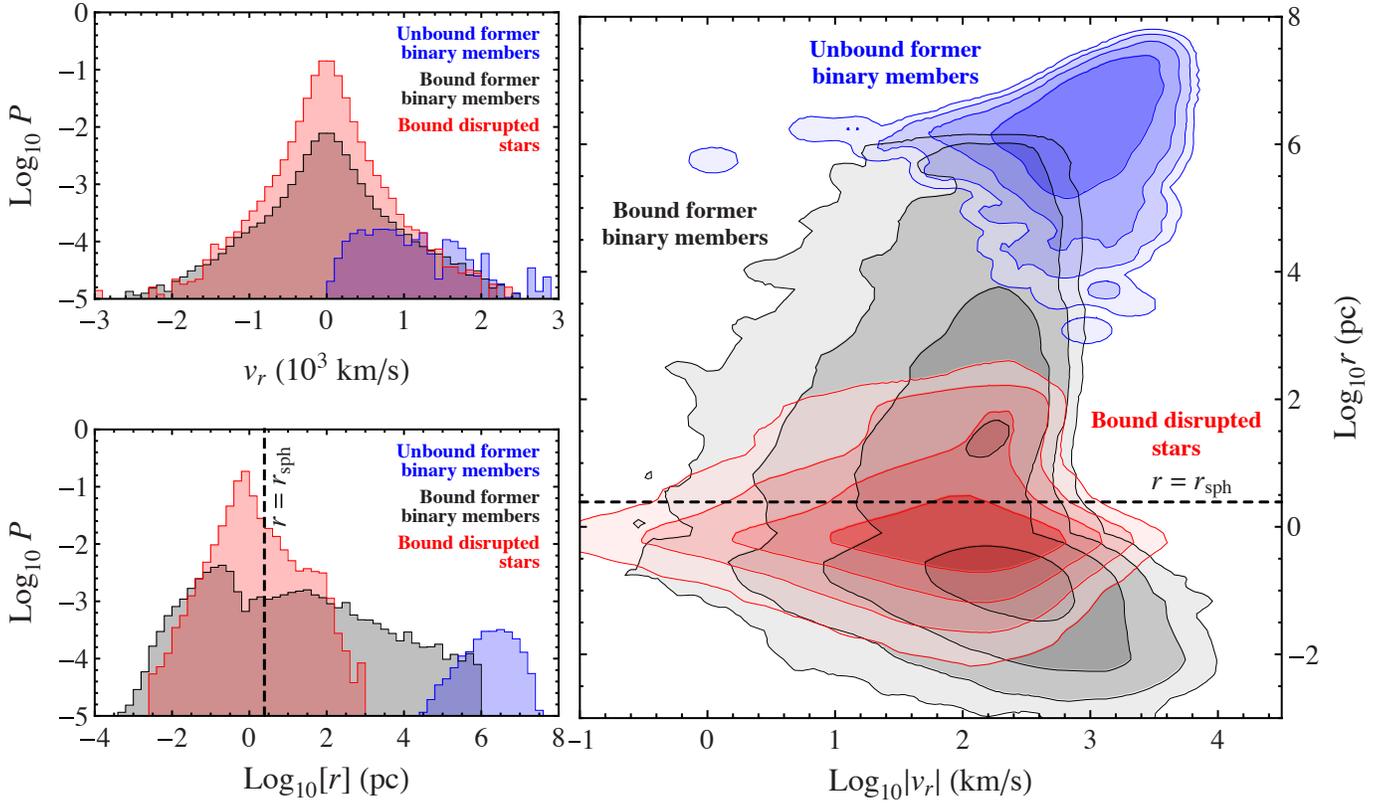}
\caption{Observed probability distributions of radial velocity $v_{r}$ and distance $r$ of the populations of stars ejected from the SMBH (generated from $4 \times 10^{4}$ encounters). Shown in blue (black) are the high-velocity (bound) members of pairs of stars that were originally members of a binary system, where the binary has had a close encounter with the SMBH, and shown in red are solitary stars that were partially disrupted by the SMBH.}
\label{fig:histograms}
\end{figure*}

Analogous to a simple harmonic oscillator, the amount of energy stored within a mode is proportional to the square of the displacement ${\cal D}$. In linear theory, the amplitudes of each mode are independent from one another, and thus the energy stored within each mode is proportional to the square of its individual displacement. By taking the square root of the two terms in Equation (\ref{eq:deleosc}) independently, we find
\begin{align}
{\cal D}_{2} &\propto \beta^{3} T_{2}^{1/2}\label{eq:d2}\\
{\cal D}_{3} &\propto q^{-1/3}\beta^{4} T_{3}^{1/2},\label{eq:d3}
\end{align}
where we have made the appropriate substitutions for $r_{\rm p}$, and where ${\cal D}_{2}$ and ${\cal D}_{3}$ are the mean displacements induced by the $l = 2$ and $l = 3$ modes. In writing Equations (\ref{eq:d2}) and (\ref{eq:d3}), we have made the assumption that the displacement due to the excitation of each mode is purely radial, but this is only true in an average sense, as the amplitude of the excitations vary over angle. In the linear approximation, the amplitude of the $l = m = 2$ and $l = m = 3$ modes are
\begin{align}
A_{2} &= R_{\ast} + {\cal D}_{2} \frac{1}{2}\sqrt{\frac{15}{2\pi}} \sin^{2} \theta \sin 2 \phi\label{eq:y22}\\
A_{3} &= R_{\ast} - {\cal D}_{3} \frac{1}{4}\sqrt{\frac{35}{\pi}} \sin^{3} \theta \sin 3 \phi\label{eq:y33}
\end{align}
where $\theta \in [0, \pi]$ is the latitude on the star's surface and $\phi \in [0, 2\pi)$ is the longitude. In writing the harmonics in this form, we set $\phi = \nu$, where $\nu$ is the true anomaly. To determine how the displacement behaves as a function of $\phi$, we integrate Equations (\ref{eq:y22}) and (\ref{eq:y33}) over $\phi$, where $0 < \phi < \pi$ corresponds to the near-side of the star, and $\pi < \phi < 2 \pi$ corresponds to the far-side. For the $l = 2$ harmonic, the near- and far-side integrals are equal and positive ($+1$), as $A_{2}$ is symmetric about $\phi = 0$, but for $l = 3$ the results of these two integrals are equal and opposite ($\pm 2/3$). Therefore, the displacement on the near-side of the star is enhanced by a factor ${\cal D}_{3}$, but reduced on the far-side of the star by $-{\cal D}_{3}$. This differential displacement is obvious when comparing two simulations with identical $\beta$ but different $q$ (Figure \ref{fig:contours}).

When a star begins to shed mass, the three pieces of the star (the surviving core, mass lost from L1, and mass lost from L2) interact with one another gravitationally. After disruption, the surviving core can exchange orbital energy with material that becomes strongly bound to the black hole, potentially enabling the core to become unbound even in cases where $\epsilon_{0} < 0$. The total change in orbital energy $\Delta \epsilon_{\rm orb}$ is a combination of $\Delta \epsilon_{\rm osc}$ and the kick applied by the two debris tails on the surviving core $\Delta \epsilon_{\rm kick}$, Equation (\ref{eq:deleosc}),
\begin{equation}
\Delta \epsilon_{\rm orb} = \Delta \epsilon_{\rm osc} + \Delta \epsilon_{\rm kick}.
\end{equation}

To estimate $\Delta \epsilon_{\rm kick}$, we calculate the specific impulse applied by the bound and unbound debris,
\begin{align}
I_{\rm L1,L2} &= a_{\rm L1,L2} t_{\rm p}\nonumber\\
&\simeq \frac{G M_{\rm L1,L2}}{R_{\ast}^{2}} \sqrt{\frac{r_{\rm p}^{3}}{G M_{\rm h}}},
\end{align}
where $t_{\rm p} \equiv \smash{(r_{\rm p}^{3}/G M_{\rm h})^{1/2}}$ is the pericenter passage time and $a_{\rm L1,L2}$ are the accelerations applied by either the bound or unbound debris tail. As the entire frame of reference travels at $v_{\rm p}$, the change in kinetic energy experienced by the surviving core is
\begin{align}
\Delta \epsilon_{\rm kick} &= \frac{1}{\sqrt{2}} \left(I_{1} - I_{2}\right) v_{\rm p}\nonumber\\
&\simeq \frac{G \Delta M_{12}}{\sqrt{2} \beta R_{\ast}} q^{1/3},\label{eq:delekick}
\end{align}
where $\Delta M_{12}$ depends on $\beta$, $q$, and $\gamma$.

The total amount of mass lost by the star as a function of $\beta$ is difficult to calculate directly from the displacement ${\cal D}$, necessitating hydrodynamical simulation \citep{RamirezRuiz:2009gw,Lodato:2009iba,Guillochon:2013jj,Cheng:2013cm}. But as the differential displacement ${\cal D}_{3}$ from the $l = 3$ mode is small compared to the size of the star, the density across ${\cal D}_{3}$ is approximately constant, and thus the differential mass loss scales directly with displacement, $\Delta M_{12} \propto {\cal D}_{3} \propto q^{-1/3}$ (Equation \ref{eq:d3}). Therefore, as the specific impulse applied by the bound and unbound debris tails is proportional to $q^{1/3}$ (Equation \ref{eq:delekick}), $\Delta \epsilon_{\rm kick}$ is independent of $q$.

\section{Discussion}\label{sec:discussion}

\subsection{Location}

As $v_{\rm kick}$ for MS stars after a disruption is maximally $v_{\rm esc} \sim $ several hundred km/s, a significant fraction of TVS are placed onto new orbits with apocenters that enclose a mass in stars $> M_{\rm h}$. By contrast, the disruption of a binary can yield $v_{\rm kick} > 4 \times 10^{3}$ km/s, and stars that receive these kicks quite easily escape the Milky Way's potential, producing hypervelocity stars \citep{Hills:1988br,Brown:2005gs} while leaving a tightly-bound companion \citep{Ghez:2005ck,Gillessen:2009fn}.

To construct the phase-space distributions of disruption products, we assume a 10\% binary fraction and a maximum binary separation of 0.5 au, drawing the primary (or solitary) mass $M_{1}$ from \citet{Kroupa:2001ki}, with $M_{1} > 0.1 M_{\odot}$, and the secondary mass assuming $M_{2} \leq M_{1}$ and $P(M_{2}) \propto M^{-0.5}$ \citep{Reggiani:2011is}. Solitary stars/binaries are assumed to initially orbit the black hole with semi-major axis $a = r_{\rm sph}/2$. We presume that $P(\beta) \propto \beta^{-2}$, $P(a_{\ast}) \propto 1/a_{\ast}$ ($a_{\ast}$ being the binary semi-major axis), $P(e)$ is given by a Rayleigh distribution with $\sigma = 0.3$, and orientation/phase of the binary's orbit are random.

In Figure~\ref{fig:histograms}, we show the distributions of the position $r$ and radial velocity $v_{r}$ that would be observed arising from the disruptions of both binary and solitary stars. Within $\Sim 100$ pc, TVS are $\Sim 10$ times more common than former binary members. One may wonder whether these stars have a significant chance of returning their original pericenter where they would likely be disrupted again (and perhaps destroyed), but the time required for stars with apocenters near $r_{\rm sph}$ on nearly-radial orbits to experience a change in angular momentum comparable to their own is on the order of their own orbital period \citep{Magorrian:1999vz}. However, the timescale for two-body relaxation in the galactic center is very long, $\Sim 10^{10}$ yr at a distance of 10 pc \citep{Merritt:2010bz}. Thus, while we expect that the orbits of TVS would be perturbed sufficiently by neighboring stars to avoid subsequent disruption, we do not expect that these stars would be shifted onto orbits that are radically different from their original orbits.

However, the relaxation time can be orders of magnitude shorter if a population of massive perturbers are present in the region immediately beyond the sphere of influence \citep{Zhao:2002fa,Perets:2006kf,Perets:2007fo}. TVS exterior to the distance corresponding to the minimum approach distance of massive perturbers would likely be scattered out of the central cluster. Therefore, the distribution of TVS can be used to constrain the distribution of massive perturbers in the galactic center.

\subsection{Appearance}
A partial disruption event is a violent process that removes the outer layers of a star, leaving a rapidly-rotating shock-heated remnant that is initially large and hot, powered by the re-accretion of material from the tidal tails \citep{Antonini:2011ia}. After re-accretion ends, the star effectively rejoins the Hayashi track (becoming redder) and contracts on a Kelvin-Helmholtz timescale ($10^5-10^7$ yr) until nuclear fusion at its core again dominates. Given a disruption rate of $10^{-4}$ yr$^{-1}$ and that $\Sim 10\%$ of disruptions produce TVS \citep{Wang:2004jy}, $\Sim 10^5$ stars (1\% of stars within 10 pc) have been harassed at some time by the SMBH, with anywhere from $1-100$ TVS undergoing Kelvin-Helmholtz contraction at any one time.

On a longer, main-sequence timescale, the increase in the mean molecular weight of the star $\mu$ that occurs due to the removal of its hydrogen-rich outer layers results in TVS being smaller, bluer, and more luminous than a MS star of the same mass and age \citep{Alexander:2001jt}.  This shift may be further enhanced by rotational mixing induced by the disruption itself or a convective stage in its subsequent contraction.

\subsection{Final Thoughts}
 
In this {\it Letter} we have shown that partially disrupted stars can receive kicks on the order of their own escape velocities, independent of the mass ratio $q$. We demonstrated that these kicks produce a population of stars within the Milky Way with $a \gtrsim r_{\rm sph}$ that have experienced a close encounter with the central black hole.

To date $\Sim 10$ stars have been identified as being unbound to the central black hole and luminous stellar cusp \citep{Reid:2007iu,Trippe:2008dj,Genzel:2010jk}. While it would be premature to claim that these particular stars are TVS without further study, TVS would appear as high-velocity interlopers within the black hole's sphere of influence, and would be brighter, at first redder, and then bluer, than stars of similar mass. And although the velocity required to escape the Milky Way's potential is larger than the typical $v_{\rm kick}$, the kicks are more than sufficient to eject stars from less-massive objects such as dwarf galaxies and globular clusters which may host their own black holes \citep{Reines:2011cs}.

\acknowledgments 
We thank F.\ Antonini, W. Brown, J.\ Faber, M.\ MacLeod, K.\ Shen, and the anonymous referee for their comments. We thank T. Bogdanovi\'{c} for spotting an algebraic mistake in our derivation of ${\cal D}$ in the originally published version.

\bibliographystyle{apj}
\bibliography{apj-jour,library}

\end{document}